# Indian Peak Power demand Forecasting : Transformer Based Implementation of Temporal Architecture

Vishvaditya Luhach†
Department of Computer Science and Engineering
Maharaja Surajmal Institute of Technology
New Delhi, India
vishvadityaluhach@gmail.com

Shashwat Jha†
Department of Electronics and Electrical Engineering
Birla Institute of Technology Mesra
Ranchi, India
shashwat6jha@gmail.com

*Abstract*— **The long-term forecasting of electricity demand has been a prevalent research topic, primarily because of its economic and strategic relevance. Several machine learning as well as deep learning techniques have been developed in parallel with the growing complexity of the peak demand, planning for generation facilities and transmission augmentation in future. Most of these proposed techniques work on short-term forecasting as long-term forecasting is considerably more challenging due to unpredictable and unforeseeable variables that may arise in the future. This paper proposes a Temporal Fusion Transformer based deep learning approach for long term forecasting of peak power demand. The dataset used in this paper consists of peak power demand in India for a period of 6 years and the prediction was done for a period of 1 year. Our proposed model was compared with other popular forecasting models and it performed considerably better in benchmarks and was also more accurate in modelling the variance in the power demand.**

## I. INTRODUCTION

Long term power demand forecasting is one of the most vigorously researched domains pertaining to its relevance in the socio-economic functioning of various countries. Peak electricity demand of a given time frame is subject to a range of uncertainties, including underlying growth in population, the development of technology, various aspects of the of the economy, changing atmospheric conditions (and the timing of those conditions), along with the variations in the usage of electricity in individual households. Peak power demand is also subject to the time, day (weekdays vs. weekends), the season and public holidays [1]. Accurate forecasting methods can be of immense value especially in the development of power supply strategy and development plan in developing countries like India where the demand is increasing at a rapid rate with high variations [2]. With the consistently increasing power demand, Judicious utilization of all relevant resources becomes of extreme importance hence contributions in the development of better forecasting techniques are pivotal.

## II. LITERATURE REVIEW

Upon exploration it's clear that power demand forecasting has been addressed by using various approaches based on machine learning as well as deep learning. Some of the early approaches during late 1990s for power forecasting include applications of regression analysis comprising of various techniques like composite multi-regression decomposition-based model [3] and multiple linear regression [4].

Further developments led to applications of stochastic time series models like ARIMA [5] and SARIMA [6] for short term as well as long term demand forecasting.

In terms of approaches based on deep neural networks, prior research shows applications of cascaded artificial neural network models [7], ANNs with quasi-newton based back propagation and utilization of principal component analysis (PCA) have also been proposed [8]. Further developments also lead to applications of recurrent neural networks (RNN) with dynamic time warping [9]. Probabilistic density-based forecasting techniques [10] as well as Relevance vector machine (RVM) based approaches [11] have also been presented especially for long term forecasting.

Prior works which have been presented do not address the extreme outliers and variations in peak power demand data of a country like India. Moreover, the trade-off between the accurate prediction, computational cost and the time taken is significant. Therefore, making it extremely important to investigate new and more efficient methods for long term peak power demand forecasting.

With this paper we intend to propose the application of transformer based TFT model and investigate its performance while comparing it to other deep learning and statistical models in accurately predicting the long-term peak power demand in India.

## III. DATASET

The Indian Power Sector Evening Hour Peak Demand data from January 2014 to the last day of 2019 within which the daily overall peak demand data has been utilized for our experiment [12] with 2191 datapoints in tabular format. The data was scaled between 0 and 1 before feeding it to the models.

---

† Equal Contribution

## IV. PROPOSED METHODOLOGY

### A. Temporal Fusion Transformer (TFT)

Introduced by Lim et al [14] temporal fusion transformer pertaining to its intricate architecture became one of the most powerful tools in forecasting. TFT inputs static metadata and future and past inputs that also vary with time. Based on these inputs, Variable Selection prudently segregates the most pertinent features. Gated Residual Network blocks [15] support dynamic flow of information through the utilization of skip connections as well as gating layers. LSTMs are used for Time-dependent processing of information locally [19], and the integration of information from any time step is done by multi-head attention [16]. Figure 1 depicts the complete in-depth architecture of TFT. The major constituents of TFT and their functions are described below:

*1) Gating mechanisms*: These are utilised to reduce extra computational cost by skipping components of the architecture which are not utilised for the task, providing adaptive depth and network complexity to be tolerant towards a variety of datasets and scenarios.

*2) Variable selection networks*: These selection networks assist in selecting relevant input variables for every time step.

*3) Static covariate encoders*: Covariate encoders are utilised for the integration of fratures which are static into the model by encoding of context vectors for the temporal dynamic health of the process.

*4) Temporal processing*: It is utilised to learn both short-term as well as long-term temporal relationships from both known as well as observed inputs that vary with time. The local processing of information is done by a sequence-to-sequence layer, whereas a multi-head attention block is used for capturing long-term dependencies .The multihead attention bloock represented in Figure 2 linearly maps the attention outputs. Fisrtly, a sequence of embeddings if fed into the encoder. 'Query' (Q), 'key' (K) and 'Value' (V) vectors are then generated by using linear transformations. The attention outputs for each embedding are then calculated using dot product of the Q, K and V vectors. Calculation of self-attention occurs multiple times independently and in parallel, hence, Multi-head Attention. The attention defines the interrelation in the patches and improves predictions.

*5) Prediction intervals via quantile forecasts*: These intervals are utilised specifically to realise the range of high probabiliy destination values.

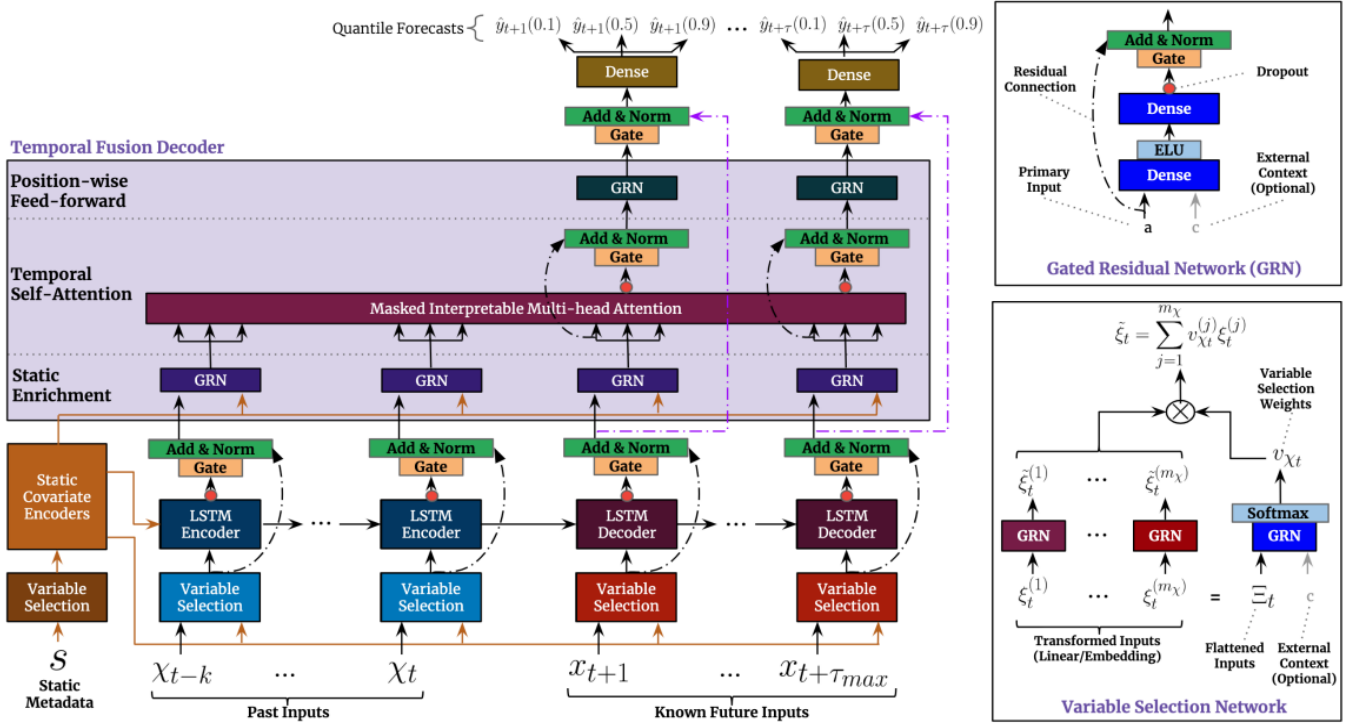

Fig. 1. Temporal Fusion Transformer (TFT) Architecture

## B. Temporal Convoltional Network (TCN)

First introduced in 2016 by Lea et al [17] temporal convolutional networks were initially proposed for video-based action segmentation. Later post identifying its high-level capability in capturing temporal relationships similar to RNNs and the training time being a fraction of what it would be with RNNs it was utilised for time series prediction as well. A great example of which was presented by Yan et al [18] where he utilised TCN as a part of his ensemble method to predict El Niño-Southern Oscillation (ENSO) in advance. TCNs contain two kinds of convolutions one is called Causal whereas the other Dilated which contribute to TCNs ability to prevent information "leakage" within a time frame and the ability to map an input sequence to an output sequence of the same length. TCNs also have very long effective history using a combination of dilated convolutions and very deep networks. Temporal Fusion Transformer Architecture

## C. Stacked LSTMs

LSTM (Long Short-Term Memory) was first introduced in 1996 by Hochreiter et al [19] whereas the full backpropagation and bi directional LSTMs were introduced by Graves.A. and Schmidhuber. J. in 2005 [20] Since then LSTMs have been utilized on multiple fronts in the domain of artificial intelligence [21-23]. It is derived from Recurrent Neural Networks (RNNs) that can efficiently learn long-term dependencies, especially in sequence prediction problems. LSTMs have feedback connections implying they are capable of processing the entire sequence of data, apart from single data points such as images. A 'Cell State' is a memory cell and plays a pivotal role in an LSTM model by maintaining its state over time. Gates control the flow of information that is fed or extracted from the memory cells by utilizing pointwise multiplication operation and sigmoid NN layers. Stacking LSTMs together has proven to be an exceptional forecasting technique since its inception and has been widely utilized for applications in various time series problems. Figure 3 depicts the architecture of a stacked LSTM model.

## D. Naïve Forecasting (Seasonal + Drift)

This method utilizes two of the models based on the Naïve forecasting methods. The seasonal forecasting model always predicts the value of *K* time steps ago. When *K=1*, this model predicts the last value of the training set. When *K>1*, it repeats the last *K* values of the training set whereas the Naïve drift model fits a line between the first and last point of the training series, and extends it in the future.

## V. EXPERIMENTATION

All the experiments were performed using TensorFlow, Keras and Darts API. All the models in this experiment were trained for 200 epochs, except for the naïve models and the performance was compared for the prediction of Indian peak power demand. Following are the system specifications on which experiments were performed: CPU - AMD Ryzen 5 3600, GPU - Nvidia 1660ti (6GB) and 16GB of system memory

The experiment was conducted such that the dataset after pre-processing was fed into 4 different models for forecasting which are explained in this section and their performance was compared on the basis of MAPE (mean absolute percentage error). Future covariates [13] were utilised for this time series prediction task. Future covariates are time series whose future values are available at prediction time. More precisely, for a prediction made at time *t* for a forecast horizon *n* and lookback window *k*, the values at times ($t+1$, ..., $t+n$) and past values for times ($t-k$, $t-k+1$, ..., $t$) of future covariates are also known.

The input chunk length for the deep learning models – TFT, LSTM, and TCN - was set at 30 and dropout rate was set at 0.1. Mean squared error was used as the loss function and a batch size of 24 was taken for all the models. For the TFT model, the output chunk length (forecast horizon) was set as 36. The number of various layers in the TFT model were – 64 hidden layer, 4 LSTM layers, and 2 Attention heads. The LSTM model had 25 hidden layers and 3 RNN layers. For the TCN model, the output chunk length was set at 28 with kernel size-3 and 6 filters with a total of 4 layers.

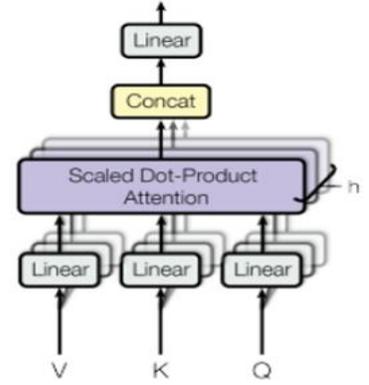

Fig. 2. Multi-Head Self Attention Block

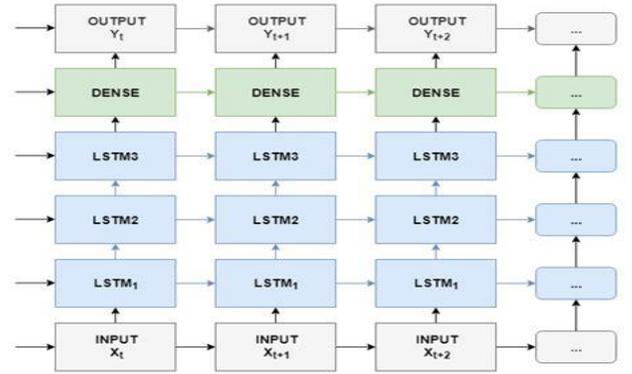

Fig. 3. Stacked LSTM Architecture

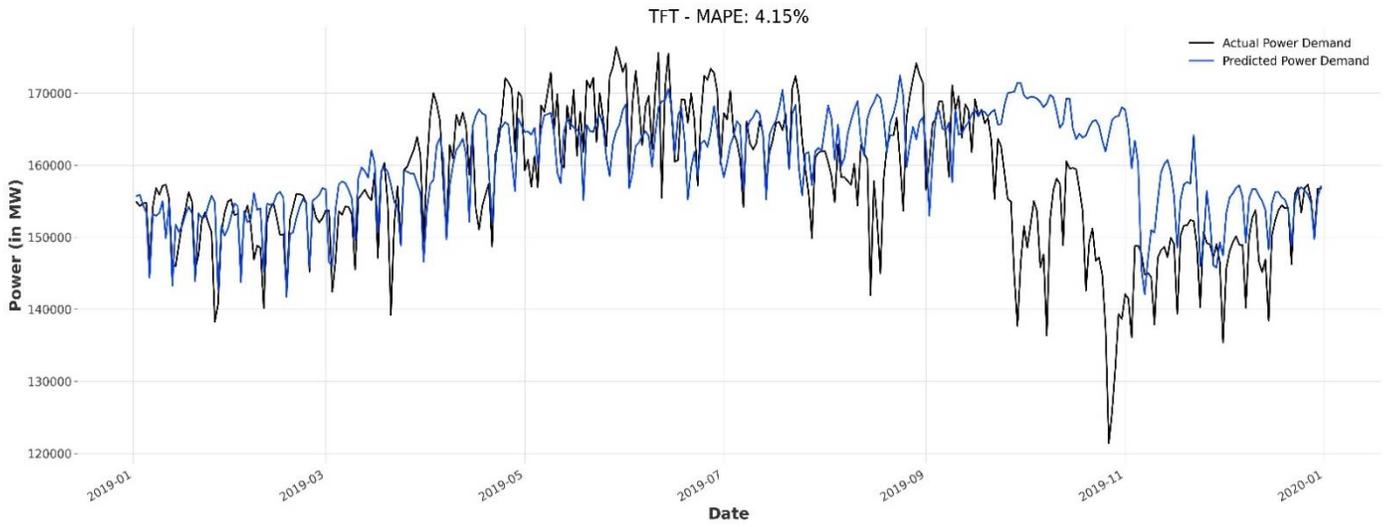

Fig. 4. Temporal Fusion Transformer Predicted Output

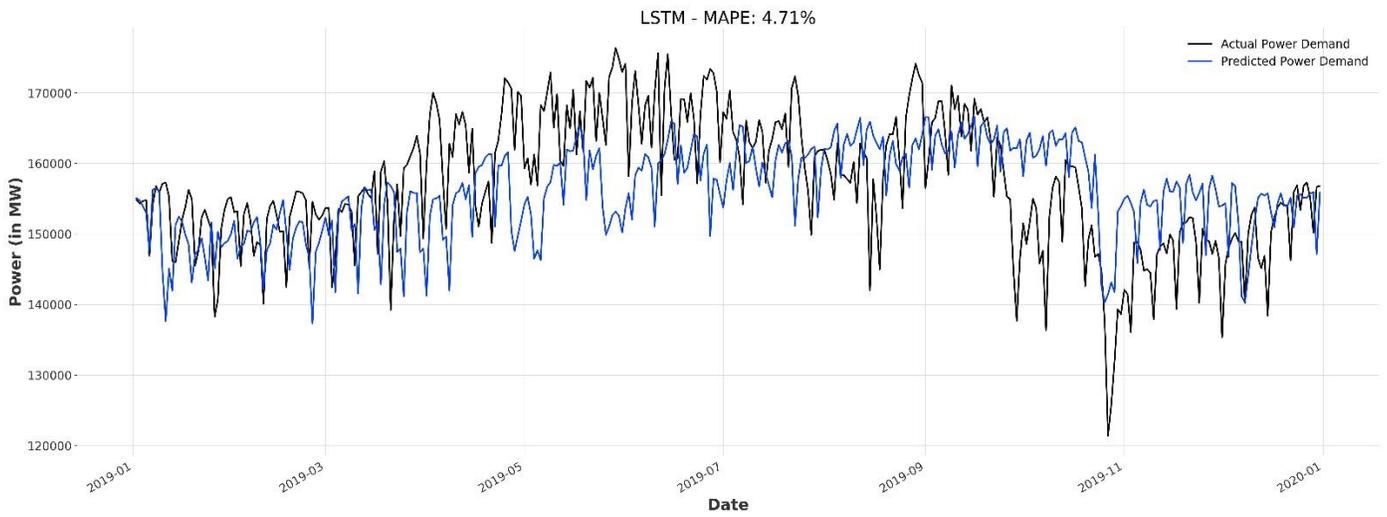

Fig. 5. LSTM Model Predicted Output

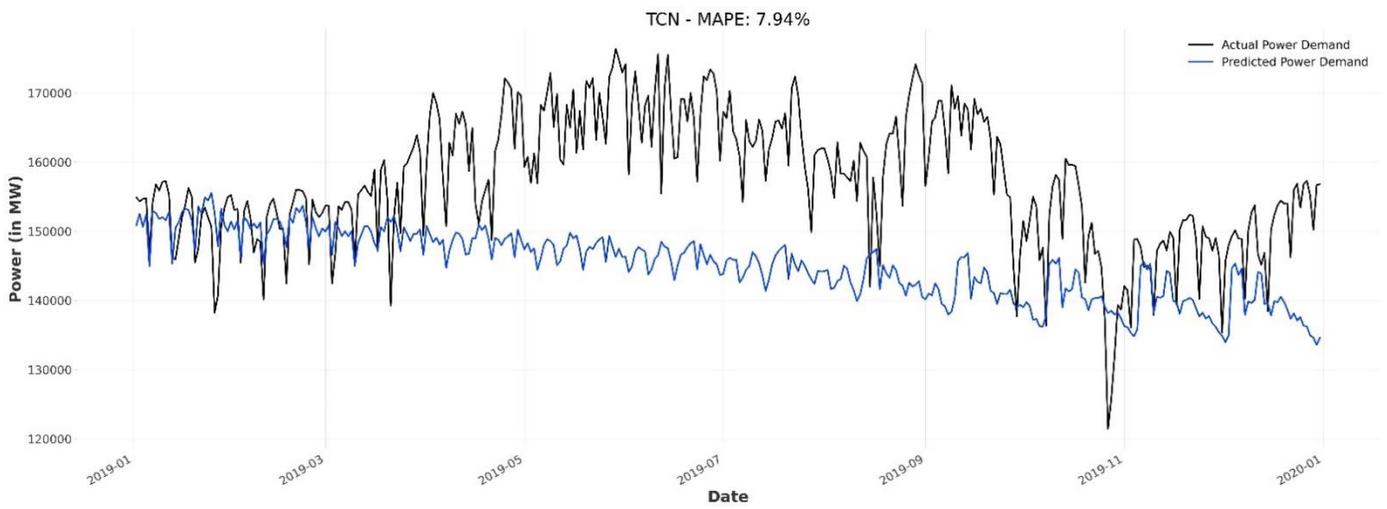

Fig. 6. Temporal Convolutional Network Predicted Output

## VI. RESULTS

Upon application of all the mentioned models the performance was analysed on the basis of MAPE (mean absolute percentage error). Cross model comparison shows that Naïve Forecasting predicted the power demand with MAPE 5.06% while Temporal Convolutional Network performed the task with MAPE 7.94%. Stacked LSTMs based technique attained MAPE 4.71% whereas the attention based TFT (Temporal Fusion Transformer) attained an MAPE of 4.15%. The cross-model performance comparison is exhibited in Table 1 whereas the figures [4-6] depict the prediction performance in the graphical format.

TABLE I - Cross model performance comparison on the baiss of M.A.P.E.

| CROSS-MODEL PERFORMANCE COMPARISON | |
|---|---|
| *Model* | *M.A.P.E.* |
| **TFT** *(ours)* | **4.15%** |
| Stacked LSTMs | 4.71% |
| TCN | 7.94% |
| Naïve Forecasting | 5.06% |

## VII. CONCLUSION

Post extensive analysis of the aforementioned models, it is evident that transformer-based models are highly effective in predicting the peak power demand in India considering the nature of the dataset. The Temporal Fusion Transformer (TFT) had the best performance amongst all followed by the stacked LSTM model. The Temporal Convolutional Network (TCN) showed the poorest performance in this task, whereas Naïve Forecasting performed well but wasn't at par with TFT and LSTMs.

This work effectively show that transformer-based models have inherently great performance and caliber when it comes to time series prediction, and future research can further improve upon this work by fine tuning the TFT architecture and utilize a higher quality dataset that can capture more variations and dependencies which affect peak power demand in a region or country.